\documentclass[prb, twocolumn,superscriptaddress, floatfix]{revtex4-1}
\usepackage{bibunits}
\usepackage{amsmath,amssymb,bm}
\usepackage{graphicx}
\usepackage{epstopdf}
\usepackage{latexsym}
\usepackage{subfigure}
\usepackage{color}
\usepackage{natbib}
\usepackage{hyperref}
\usepackage{braket}
\hypersetup{
  colorlinks,
  citecolor=magenta,
  linkcolor=blue,
  urlcolor=blue}

\begin{document}

\title{First-order N\'eel-cVBS transition in a model square lattice $S=1$ antiferromagnet}
\author{Julia Wildeboer}
\affiliation{Department of Physics \& Astronomy, University of Kentucky, Lexington, KY-40506-0055}
\author{Nisheeta Desai}
\affiliation{Department of Physics \& Astronomy, University of Kentucky, Lexington, KY-40506-0055}
\author{Jonathan D'Emidio}
\affiliation{Institute of Physics, \'{E}cole Polytechnique F\'{e}d\'{e}rale de Lausanne (EPFL), CH-1015 Lausanne, Switzerland}
\author{Ribhu K. Kaul}
\affiliation{Department of Physics \& Astronomy, University of Kentucky, Lexington, KY-40506-0055}

\begin{abstract}
We study the N\'eel to four-fold columnar valence bond solid (cVBS) quantum phase
transition in a sign free $S=1$ square lattice model. This is the same kind of transition that for $S=1/2$ has been argued to realize the prototypical deconfined critical point. Extensive numerical simulations of the square lattice $S=1/2$ N\'eel-VBS transition have found consistency with the DCP scenario with no direct evidence for first order behavior. In contrast to the $S=1/2$ case, in our quantum Monte Carlo simulations for the $S=1$ model, we present unambiguous evidence for a direct conventional first-order quantum phase transition. Classic signs for a first order transition demonstrating co-existence including double peaked histograms and switching behavior are observed. The sharp contrast from the $S=1/2$ case is remarkable, and is a striking
demonstration of the role of the size of the quantum spin in the phase diagram of two dimensional lattice models. 
\end{abstract}

\maketitle

\section{Introduction}

The destruction of N\'eel order by quantum fluctuations is a hotly studied
issue in quantum magnetism inspired originally by the parent compounds of cuprate high temperature
superconductors. In the cuprates, the N\'eel order appears for $S=1/2$ moments on the square lattice. In this case, many theoretical arguments
and extensive unbiased numerical calculations have put forth evidence for a four-fold
degenerate columnar valence bond
solid (VBS) phase on the destruction of N\'eel order, possibly separated by the novel
deconfined critical
point.~\cite{read1989:vbs,senthil2004:science,senthil2004:deconf_long,sandvik2007:deconf,melko2008:jq} More recently, inspired
by the iron pnictide superconductors,
a number of
studies of the destruction of N\'eel order in $S=1$ square lattice
systems have appeared,~\cite{wang2015:s1,yu2015:s1,hu2017:nem,corboz2017:hn}
building on previous studies of the phase diagram of square lattice
$S=1$ systems, (see
\cite{toth2012:s1,jiang2009:s1,chen2018:s1,harada2007:deconf,michaud2013:s1}
and references therein).
It is thus interesting to extend the success of unbiased
quantum Monte Carlo (QMC) studies of the destruction of N\'eel order
in square lattice $S=1/2$ systems~\cite{kaul2013:qmc} to the $S=1$
case. In previous QMC studies the phase transitions in coupled $S=1$ chains~\cite{harada2007:deconf} and the transition to the Haldane nematic have been considered.~\cite{desai2019:spsdes} Here we will study the transition between the N\'eel state and a columnar Valence bond solid. A cartoon wavefunction for such a cVBS can be simply written down since two $S=1$ spins can form a singlet from the elementary rules of the addition of angular momentum; these singlets can then be arranged in the standard columnar pattern.

The role of the microscopic value of spin on the phase diagrams of one-dimensional
spin chains is now well established. Most famously Heisenberg models
with integer spins realize a ground state with a gap to all excitations called the Haldane gap,
whereas half integer spin chains realize an interesting gapless ground
state described at long distances by the SU(2)$_1$ Wess-Zumino-Witten
field theory.~\cite{affleck1989:lh} It is interesting to ask what the role of the size of
the spin is in two dimensions? While the square lattice Heisenberg
model is N\'eel ordered for all spin-$S$, the nature of the accessible non-magnetic phases and
the theory of critical phenomena at the destruction of N\'eel order
has been argued to depend sensitively on the value of the spin.~\cite{read1989:vbs} 
Since the subtle
quantum effects that arise from Berry phase terms depend crucially the microscopic
value of the spin,~\cite{haldane1988:berry} one can
expect striking differences between $S=1/2$ and $S=1$ even for phase
transitions that appear identical with respect to the Landau-Ginzburg-Wilson criteria of
dimensionality, symmetry and order parameters.  We will study this
interesting issue here by focusing on the square lattice N\'eel-cVBS phase transition in $S=1$
magnets. The identical phase transition for $S=1/2$ is described by
deconfined criticality which has argued for a single continuous phase transition.

\begin{figure}[!t]
\centerline{\includegraphics[angle=0,width=1.0\columnwidth]{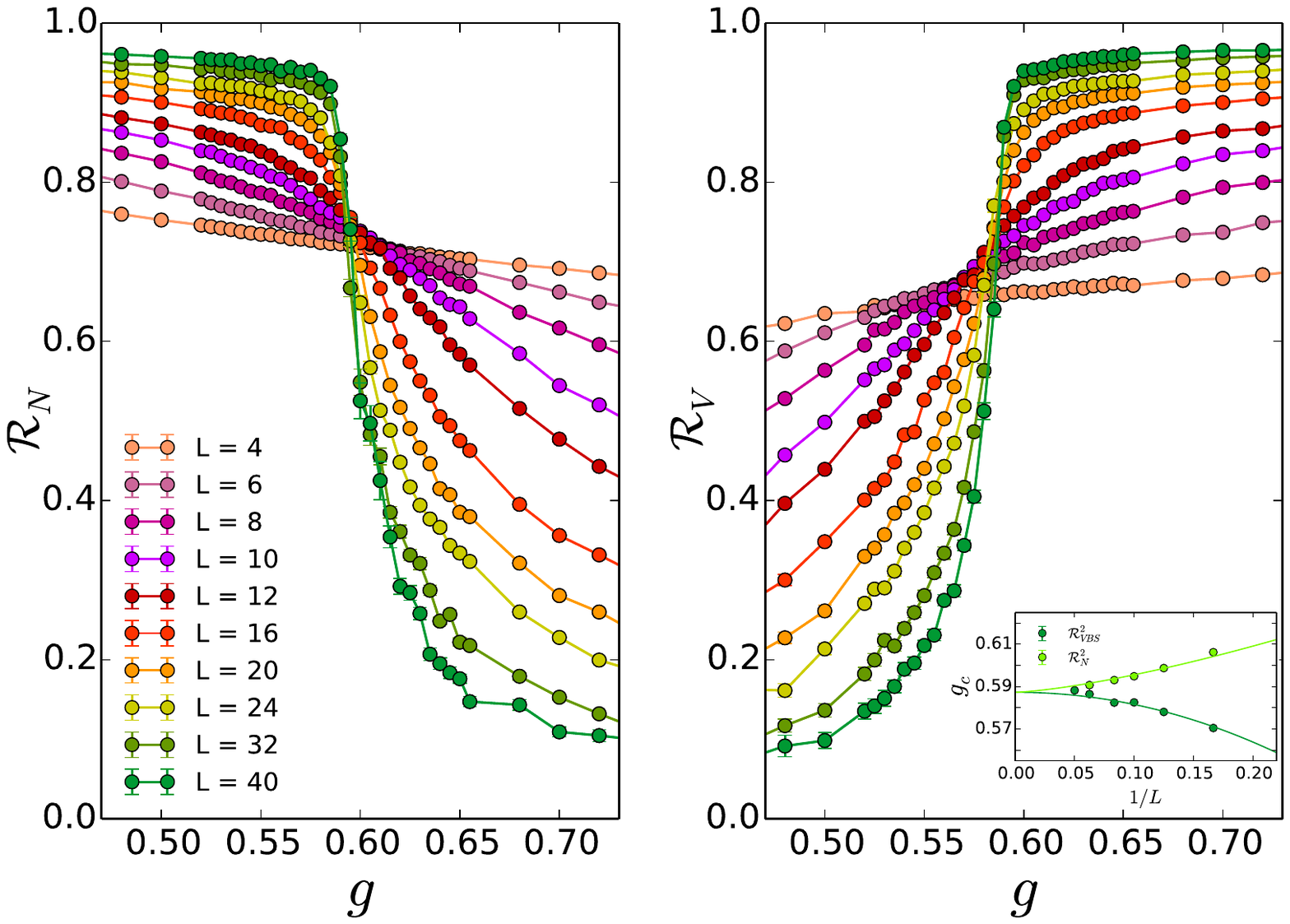}}
\caption{\label{fig:RNV} N\'eel and VBS order parameters ratios, ${\cal
    R}_N$ and ${\cal R}_V$ close to the
  quantum phase transition showing clear evidence for a direct transition. (inset) shows the
  value of $g_c$ obtained by analyzing crossings of $L$ and $2L$
  values for both ratios. solid lines are a fit to the data giving $g_c=0.588(2)$.}
\end{figure}

We note that a field theoretical study has taken up a related issue previously.~\cite{grover2007:deconf} Extending their results in a straightforward manner would suggest that a $S=1$ N\'eel-cVBS transition could possibly be described by an anisotropic CP$^2$ field theory with quadrupled monopoles. That this implies a {\em continuous} deconfined transition in our microscopic model requires a litany of additional assumptions, including that the field theory has an anisotropic fixed point, quadrupled monopoles are irrelevant at this fixed point and that our microscopic model crosses the critical surface so we can flow into the fixed point. As we shall see below in our microscopic model we find a first order transition, but it is unclear yet which of these assumptions fails. Further work on both microscopic models and field theory could shed light on this subtle detail in the future.

\begin{figure}[!t]
\centerline{\includegraphics[angle=0,width=1.0\columnwidth]{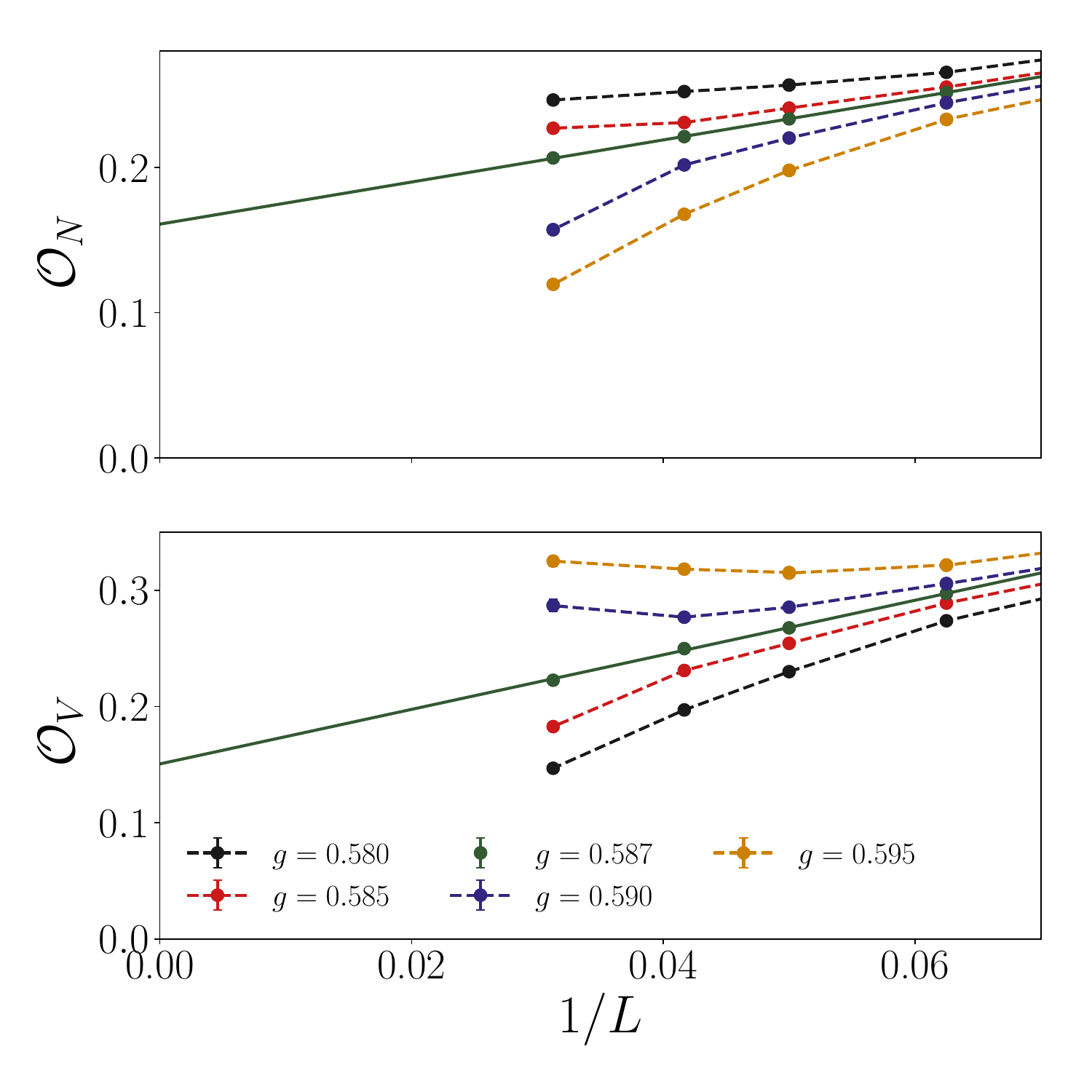}}
\caption{\label{fig:orderNV} Finite size scaling of the order parameters
  ${\cal O}_N$ and ${\cal O}_{V}$ close to the phase transition on system sizes up to $L=32$.
The extrapolations to a finite value for both N\'eel and VBS order
parameters at a common coupling $g=0.587$ point to that fact that both order parameters are finite at the transition. The dashed lines are a guide to the eye, connecting data at the same coupling value. We have used the form $\mathcal{O}_{N,V}(L)=C_0+\frac{C_1}{L}$ for the extrapolation. While the extrapolations are not expected to be quantitatively reliable, they clearly suggest that both order parameters are finite at the phase transition. Although this evidence is suggestive of co-existence and first order behavior, we present extensive evidence in Figs.~\ref{fig:ordrprmhist}-\ref{fig:switch}, which unequivocally confirms this interpretation.}
\end{figure}

\section{Model}

Our first goal is to design a $S=1$ sign free model in which
the N\'eel-cVBS transition can be studied using Monte-Carlo simulations. We start with the square
lattice $S=1$  Heisenberg model,
\begin{equation}
\label{eq:j}
H_J =  J \sum_{\langle ij \rangle}\vec{S_i} \cdot \vec{S_j}
\end{equation}
This model is well known to be N\'eel ordered.
Because we are working with $S=1$, it is possible
to square the bilinear operator and obtain an independent ``biquadratic
operator,'' $\left ( \vec S_i \cdot \vec S_j\right )^2$, also amenable to QMC.~\cite{harada2002:biq,kaul2012:biq} Using this term we can
construct a Sandvik-like four spin interaction,~\cite{sandvik2007:deconf}
\begin{equation}
\label{eq:qk}
H_{Q_K} =  -Q_K  \sum_{ijkl \in \square}\left (\left (\vec{S_i} \cdot
    \vec{S_j}\right )^2-1\right )  \left (\left (\vec{S_k} \cdot
    \vec{S_l}\right )^2-1\right )
\end{equation}
We note that $H_{Q_K}$
has a higher staggered SU(3) symmetry because it is constructed from the biquadratic
interaction, of which the physical SU(2) is a subgroup. However the model we study here $H_{JQ_K}=H_J+H_{Q_K}$
has only the generic SU(2) symmetry obtained by rotating the $\vec S$
vector in the usual way. Previous numerical studies have
established that $H_{Q_K}$ on the square lattice has four-fold
columnar VBS order.~\cite{lou2009:sun,kaul2011:su34,banerjee2010:su3} Thus the single tuning
parameter in $H_{JQ_K}$ gives us 
unbiased numerical access to the N\'eel-VBS transition in a $S=1$ system, as desired.

\begin{figure}
    \centering
    \includegraphics[width=1.0\columnwidth]{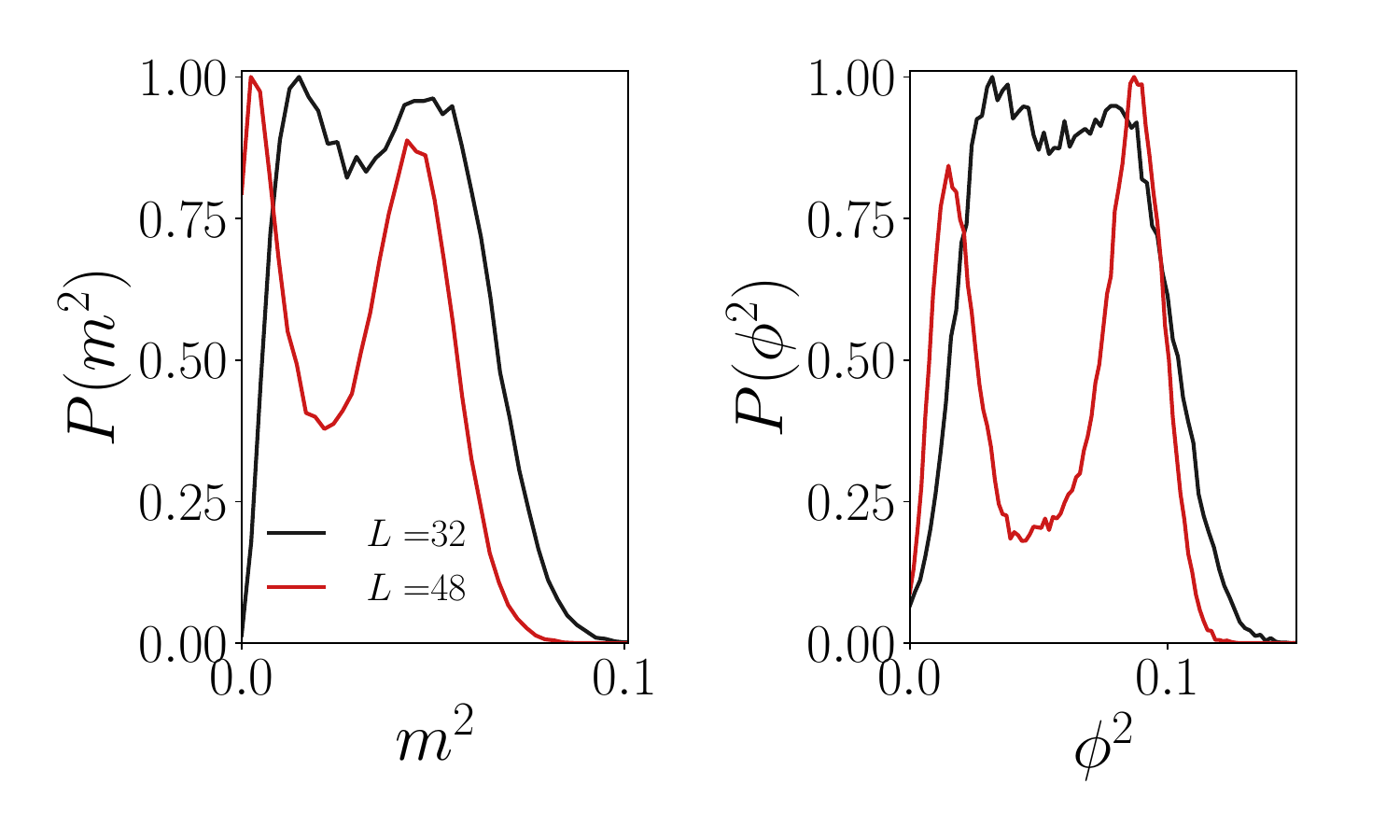}
    \caption{Histograms of our Monte Carlo estimators $m^2$ and $\phi^2$ that over the whole Monte Carlo run average to $\mathcal{O}^2_{N}$ and $\mathcal{O}^2_{V}$ respectively. Close to the transition (at $g \approx 0.588$ and $\beta=L/4$) the probability distributions of these quantities show two peaks: One of the peaks that is close to 0 corresponds to disorder and the other one at a finite value corresponds to the ordered phase. This double peak feature gets sharper as we increase system size which is evidence in support of a first order transition between the two orders.}
    \label{fig:ordrprmhist}
\end{figure}

Since our model is constructed to be Marshall sign positive, it can be
simulated without a sign problem using the stochastic series expansion
method (SSE) ~\cite{sandvik2010:vietri}. We have used two different algorithms as described in Sec. \ref{subsec:algo} which produce the same results within errors. Our simulations are carried out on $L\times L$ square lattices at an inverse temperature $\beta$ -- all the data presented here has been checked to be in the $T=0$ limit as shown in \ref{subsec:gsconv}. We work in units in which $J=1$, and define the tuning parameter $g\equiv Q_K/J$ to access the phase transition. We study  the Fourier transform of the N\'eel and VBS correlation functions, $S^N_{\bf k}=\frac{1}{L^2}\sum_r e^{i {\bf k \cdot  r}}\langle S^z({\bf r}) S^z({\bf 0})\rangle$ and $S^V_{\bf k}=\frac{1}{L^2}\sum_r e^{i {\bf k \cdot r}}\langle S(\bf r) \cdot S(\bf r + \hat{\bf x})  S(\bf 0) \cdot S(\bf 0 + \hat{\bf x})\rangle$. We define the order parameters as
${\cal O}^2_N = S^N_{\bf (\pi,\pi) }$ and ${\cal O}^2_V = S^V_{\bf
  (\pi,0) }$. For each of the order parameters we define ratios $R = 1-\frac{S_{{\bf K}+\frac{2\pi}{L}{\bf
        y}} }{S_{\bf K}}$ (with ${\bf K}$ the ordering momentum); $R$ goes to 1 in
a phase with long range order and 0 in a disordered phase.
In the SSE method we map the quantum partition function of our model to a classical loop model in one higher dimension.\cite{sandvik2010:vietri} The winding number of these loops is also a useful quantity to detect the magnetic phase. The spin stiffness defined as Eq. \ref{eq:stiff} is related to the square of the winding number of these loops, $\langle \mathcal{W}^2 \rangle$, as shown in Eq. \ref{eq:winding}. The magnetic phase is characterized by long loops with $\langle \mathcal{W}^2 \rangle$ diverging linearly with $L$ while the VBS phase has short loops with $\langle \mathcal{W}^2 \rangle$ going to zero.

\section{Numerical Results}

Fig.~\ref{fig:RNV} shows the ratios $R$ for the N\'eel and VBS order parameters as a
function of $g$ for different $L$. The data (see inset for finite size
scaling)
provides strong evidence that the N\'eel-VBS transition is direct with a $g_c=0.588(2)$ --
we can safely rule out
co-existence or an intermediate phase. We note that this study does not by itself indicate whether the transition is first order or continuous.

The ratio data leaves open the
possibility of a direct continuous transition. The first indication that this does not occur is shown in Fig.~\ref{fig:orderNV}. In this finite size
scaling plot of both order parameters, we have reasonable evidence that
at the transition both order parameters are {\em finite}. We have carried out extrapolations on system sizes up to $L=32$. While it is not fully reliable quantitatively to extrapolate the order parameter data  with such a limited system size range,  there is little doubt that both N\'eel and cVBS order parameters are finite at $g=0.587$. This would indicate a first order transition or a co-existence between N\'eel and cVBS phases.

\begin{figure}
    \centering
    \includegraphics[width=1.0\columnwidth]{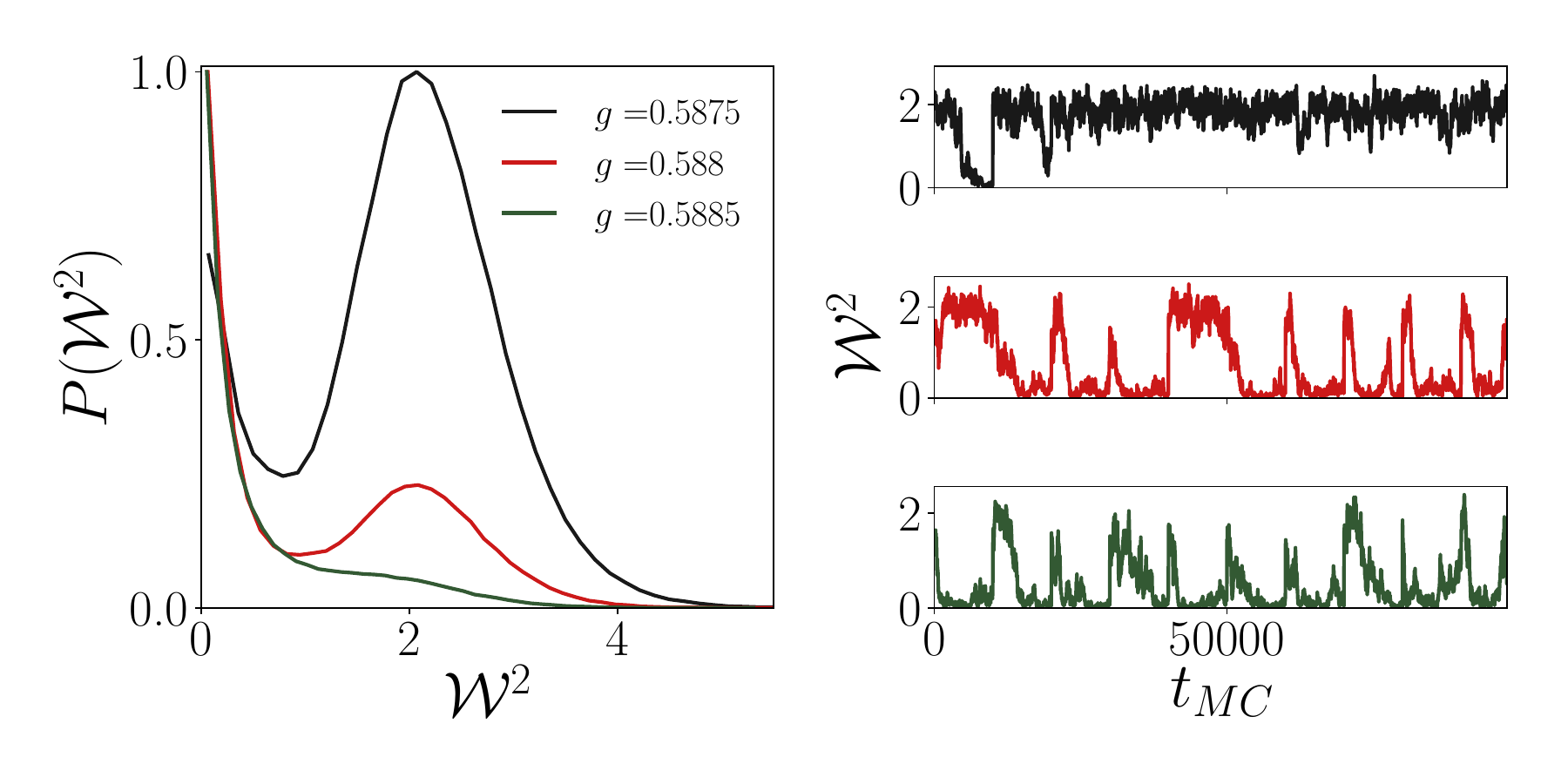}
    \caption{Histograms (left) and Monte Carlo histories (right) of $\mathcal{W}^2$ for $L=48$ and $\beta=12$. Two peaks in probability distribution of $\mathcal{W}^2$ near the critical point and switching of this quantity between zero and a finite value as a function of Monte Carlo time both point to first order behavior.}
    \label{fig:w2hist}
\end{figure}

\begin{figure}
    \centering
    \includegraphics[width=1.0\columnwidth]{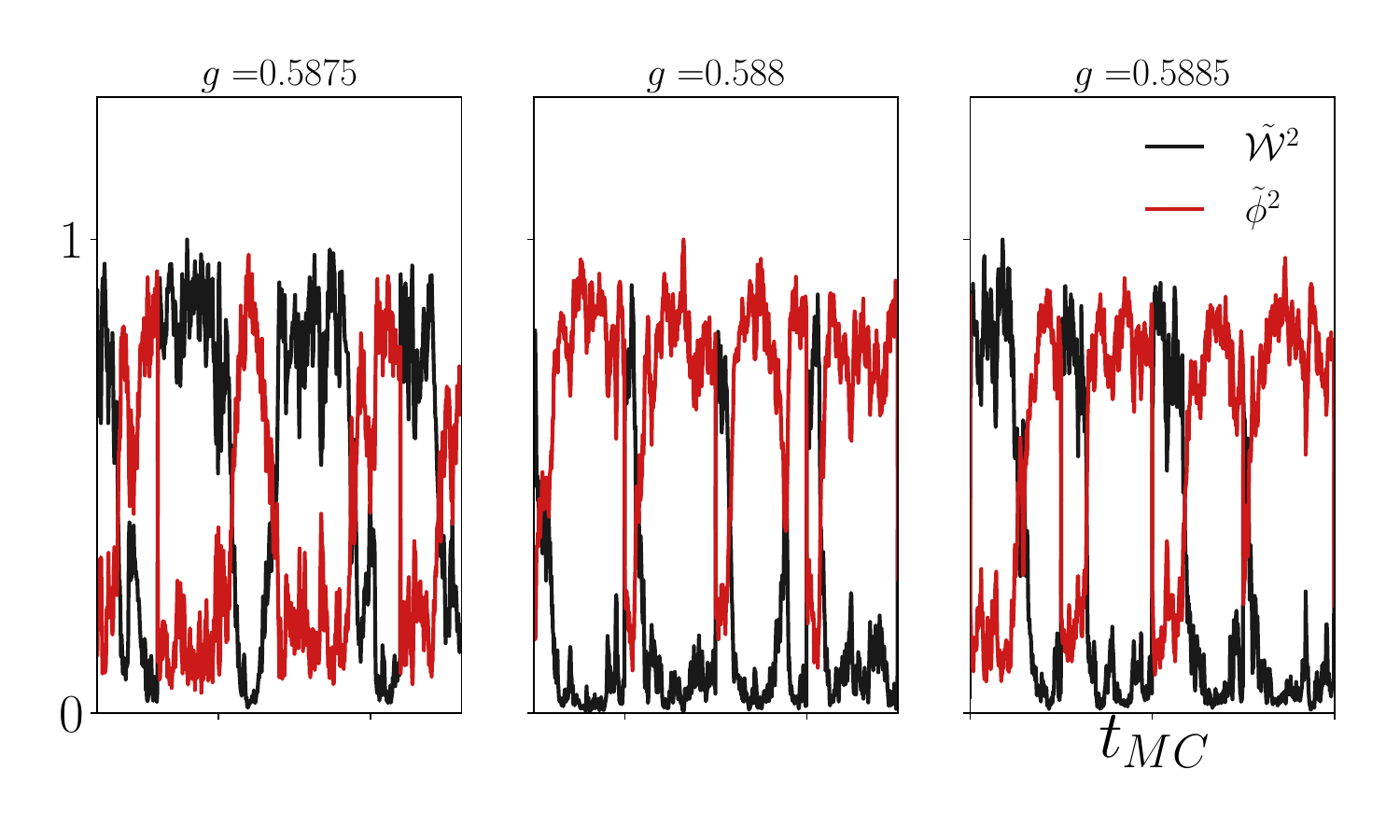}
    \caption{MC histories of $\mathcal{W}^2$ and $\phi^2$ for $L=48$ at $\beta=12$ shows clear switching behavior in both quantities at three different couplings close to the critical point (the exact couplings are shown above each of the three figures). Here $\tilde{\mathcal{W}}^2$ and $\tilde{\phi}^2$ are normalized values of $\mathcal{W}^2$ and $\phi^2$ such that the maximum is unity. It can be clearly seen that one order is present when the other is absent. We thus conclude that the system switches between the two orders at the critical point which is characteristic of a first order transition.}
    \label{fig:switch}
\end{figure}

Beyond system sizes of $L\approx 32$, it is very difficult to get QMC data with small error bars close to the critical point. As we now elaborate the reason for this is that we are encountering a first order N\'eel-cVBS transition. Fig.~\ref{fig:ordrprmhist} shows histograms for the N\'eel and cVBS order parameter estimators which show clear double peaked behavior that gets pronounced as the system size is increased. The stiffness, which is finite in the N\'eel phase and goes to zero in the cVBS phase also shows clear double peaked behavior close to the transition. The double peaked behavior results from the system switching between N\'eel and cVBS phases during the simulation. This is shown in Fig.~\ref{fig:switch} in which we observe clearly that when the magnetic order is present, the VBS order is absent and vice versa. This switching takes place as a function of Monte Carlo time indicating metastability, co-existence of the two orders and hence a first order transition.

\section{Conclusions}

We have introduced a model for the transition from the N\'eel to the four fold degenerate columnar valence bond solid state which is amenable to sign free quantum Monte Carlo simulations. Previous field theoretic work extending the $S=1/2$ deconfined criticality scenario to $S=1$ has argued that this transition could be direct and continuous, and described by an anisotropic CP$^2$ field theory. Instead, a detailed numerical study of our model shows that this phase transition is direct but of first order in our model. With no known model that shows a continuous transition it is possible that one of the assumptions of the field theoretic scenario is itself incorrect, e.g. the existence of an anisotropic SU(3) fixed point. Clearly more field theoretic work is needed to further our understanding of these interesting issues. In future numerical work it will be interesting to understand how our $S=1$ model connects to the special SU(3) point where a continuous transition has been observed in QMC simulations. Also interesting, would be to understand whether the N\'eel-cVBS transition for $S=3/2$ resembles the findings of the $S=1/2$ case as expected from field theoretic scenarios.

We acknowledge partial
financial support from NSF-DMR 1611161. We are grateful to the hospitality of the Aspen
Center for Physics (NSF grant no. 1607611). Computing
resources were obtained through NSF's XSEDE award TG-DMR-140061 and the DLX computer
at the University of Kentucky.

\appendix

\section{Numerical Details}

\subsection{Algorithm}
\label{subsec:algo}
The numerical results presented in this work have been obtained using two different methods, both of which are some adaptation of the standard Stochastic Series Expansion (SSE)\cite{sandvik2010:vietri} algorithm: 

\begin{enumerate}
    \item In the first method we work in the $S_z =-1,0,1$ basis for our $S=1$ problem. To update the SSE configurations we use both local diagonal updates and  the non-local directed loop algorithm~\cite{syljuasen2002:dirloop} that allows us to switch between the allowed vertices while respecting the $S_z$ conservation.
    \item In the second method we use the split spin representation \cite{todo2001:highs,kawashima1994:highs} where each $S=1$ is replaced by two $S=\frac{1}{2}$'s. We then simulate a $S=\frac{1}{2}$ model instead of a $S=1$ model and project out states that only belong to the $S=1$ subspace.\cite{desai2019:spsdes}  
\end{enumerate}

\subsection{ Measurements and QMC-ED comparison:}
\label{subsec:EDchecks}
We have tested our code by performing comparisons against exact
diagonalization. For future reference, Tables \ref{table0} and \ref{table1} provide test comparisons between measurements obtained from a SSE study and exact diagonalization (ED) on a lattice of size $(L_{x},L_{y}) = (4,4)$, for various combinations
of the bond and plaquette interactions $J$ and $Q_{K}$ for the
$J-Q_{K}$ model under investigation in this work and for various
combinations of the bond and plaquette interactions $J$ and $Q_{J}$
for the spin$-1$ version of Sandvik's $J-Q_{J}$ model (described in \ref{subsec:jqmodel}).  Due to the very large Hilbert space for this spin-1 model on a
4x4 lattice, we project out the ground state from a random state in the $S^z=0$ subspace, thus avoiding the need to diagonalize the sparse Hamiltonian matrix.
We list values for the extensive ground state energy, the N\'eel order parameter $\mathcal{O}^{2}_{N}$ as well as the VBS order parameter 
$\mathcal{O}^{2}_{V}$. Also shown are the so-called ratios
$\mathcal{R}_{N}$ and $\mathcal{R}_{V}$. These quantities measured using both the algorithms described in \ref{subsec:algo} have been checked to match. All observables are defined below.


\begin{figure}
    \centering
    \includegraphics[width=1.0\columnwidth]{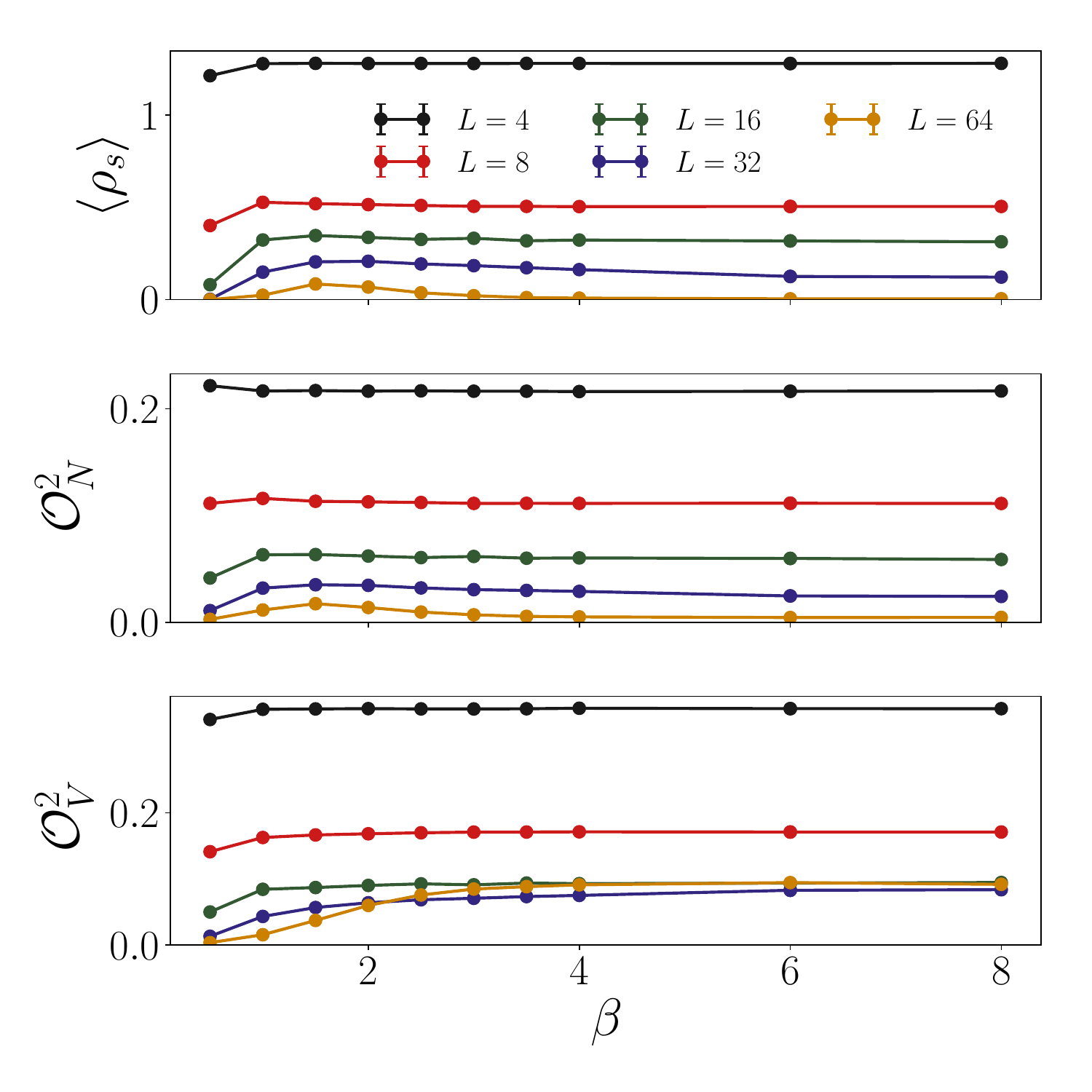}
    \caption{All observables saturate as a function of inverse temperature ($\beta$) before $\beta = 6$ at $g=0.59$}
    \label{fig:obsvsbeta}
\end{figure}

\begin{table*}[t]
    \begin{tabular}{ | c | c | c | c || c | c || c | c || c | c || c | c || c | c |}
    \hline
    $L_{x}$    & $L_{y}$       & $J$ & $Q_{K}$ & $E$ (ED) & $E$ (MC)   & $\mathcal{O}^{2}_{N}$ (ED) & $\mathcal{O}^{2}_{N}$ (MC) & $\mathcal{O}^{2}_{V}$ (ED) & $\mathcal{O}^{2}_{V}$ (MC) & $\mathcal{R}_{N}$ (ED) & $\mathcal{R}_{N}$ (MC) & $\mathcal{R}_{V}$ (ED) & $\mathcal{R}_{V}$ (MC)\\ \hline \hline
    4 &      4 &   0.2  &   0.9     &   -96.15381  &  -96.147(8) &  0.13590   & 0.13592(2)  &  0.50414   & 0.5044(5)  &   0.49940 & 0.4993(1) & 0.75713   & 0.7570(7)\\ \hline
    4 &      4 &   0.5  &   0.2     &   -49.02200  &  -49.024(4) &  0.25596   & 0.25594(9)  &  0.28370   & 0.2838(2)  &   0.78679 & 0.7868(1) & 0.59012   & 0.5907(7)\\ \hline
    4 &      4 &   0.7  &   0.3     &   -70.29052  &  -70.288(5) &  0.24879   & 0.24867(8)  &  0.29726   & 0.2971(2)  &   0.77611 & 0.7760(1) & 0.60493   & 0.6054(6)\\ \hline
    4 &      4 &   0.8  &   0.4     &   -85.17819  &  -85.180(6) &  0.23283   & 0.23291(6)  &  0.32728   & 0.3269(2)  &   0.75040 & 0.7503(1) & 0.63436   & 0.6346(5)\\ \hline
    4 &      4 &   0.9  &   0.6     &  -109.00470  & -109.001(7) &  0.20556   & 0.20562(3)  &  0.37805   & 0.3777(2)  &   0.69897 & 0.6989(1) & 0.67619   & 0.6761(4)\\ \hline
\end{tabular}
\caption{The table shows the extensive energy ($E$), the N\'eel order parameter $\mathcal{O}^{2}_{N}$ and the VBS order parameter $\mathcal{O}^{2}_{V}$
obtained by exact diagonalization (ED) and by Stochastic Series Expansion Monte Carlo (SSE) for the spin$-1$ $J-Q_{K}$ model. Additionally shown are the ratios $\mathcal{R}_{N}$ and $\mathcal{R}_{V}$. For the SSE, errors are also shown. The MC data is computed with $\beta = 40$.
}

\label{table0}
\end{table*}
\begin{table*}[tb]
\begin{tabular}{ | c | c | c | c || c | c || c | c || c | c || c | c || c | c |}
\hline
$L_{x}$    & $L_{y}$       & $J$ & $Q_{J}$ & $E$ (ED) & $E$ (MC)   & $\mathcal{O}^{2}_{N}$ (ED) & $\mathcal{O}^{2}_{N}$ (MC) & $\mathcal{O}^{2}_{V}$ (ED) & $\mathcal{O}^{2}_{V}$ (MC) & $\mathcal{R}_{N}$ (ED) & $\mathcal{R}_{N}$ (MC) & $\mathcal{R}_{V}$ (ED) & $\mathcal{R}_{V}$ (MC)\\ \hline \hline
4 &  4 &  0.2  &  0.9     &  -157.24324  &  -157.251(8) &  0.33323      & 0.3330(1) & 0.12077  & 0.121(1)  & 0.87616   & 0.87610(8) &  0.29722 & 0.295(9) \\ \hline
4 &  4 &  0.5  &  0.2     &   -66.86936  &   -66.861(3) &  0.34103      & 0.3409(2) & 0.10760  & 0.1073(3) & 0.88539   & 0.8854(1)  &  0.21940 & 0.215(3) \\ \hline
4 &  4 &  0.7  &  0.3     &   -96.79576  &   -96.790(5) &  0.34071      & 0.3406(2) & 0.10814  & 0.1079(3) & 0.88501   & 0.8850(1)  &  0.22295 & 0.225(3) \\ \hline
4 &  4 &  0.8  &  0.4     &  -119.70732  &  -119.707(4) &  0.34001      & 0.3402(1) & 0.10935  & 0.1090(2) & 0.88417   & 0.8843(1)  &  0.23071 & 0.226(3) \\ \hline
4 &  4 &  0.9  &  0.6     &  -158.52300  &  -158.520(6) &  0.33873      & 0.3388(1) & 0.11153  & 0.1113(3) & 0.88264   & 0.88268(9) &  0.24436 & 0.241(3) \\ \hline
\end{tabular}
\caption{The table shows the extensive energy ($E$), the N\'eel order parameter $\mathcal{O}^{2}_{N}$ and the VBS order parameter $\mathcal{O}^{2}_{V}$ obtained by exact diagonalization (ED) and by Stochastic Series Expansion Monte Carlo (SSE) for the spin$-1$ $J-Q_{J}$ (Sandvik's) model. Also shown are the ratios $\mathcal{R}_{N}$ and $\mathcal{R}_{V}$. For the MC, errors are also shown. The MC data is again computed with $\beta = 40$.
}
\label{table1}
\end{table*}

\begin{figure}[t]
\includegraphics[width=1.0\columnwidth]{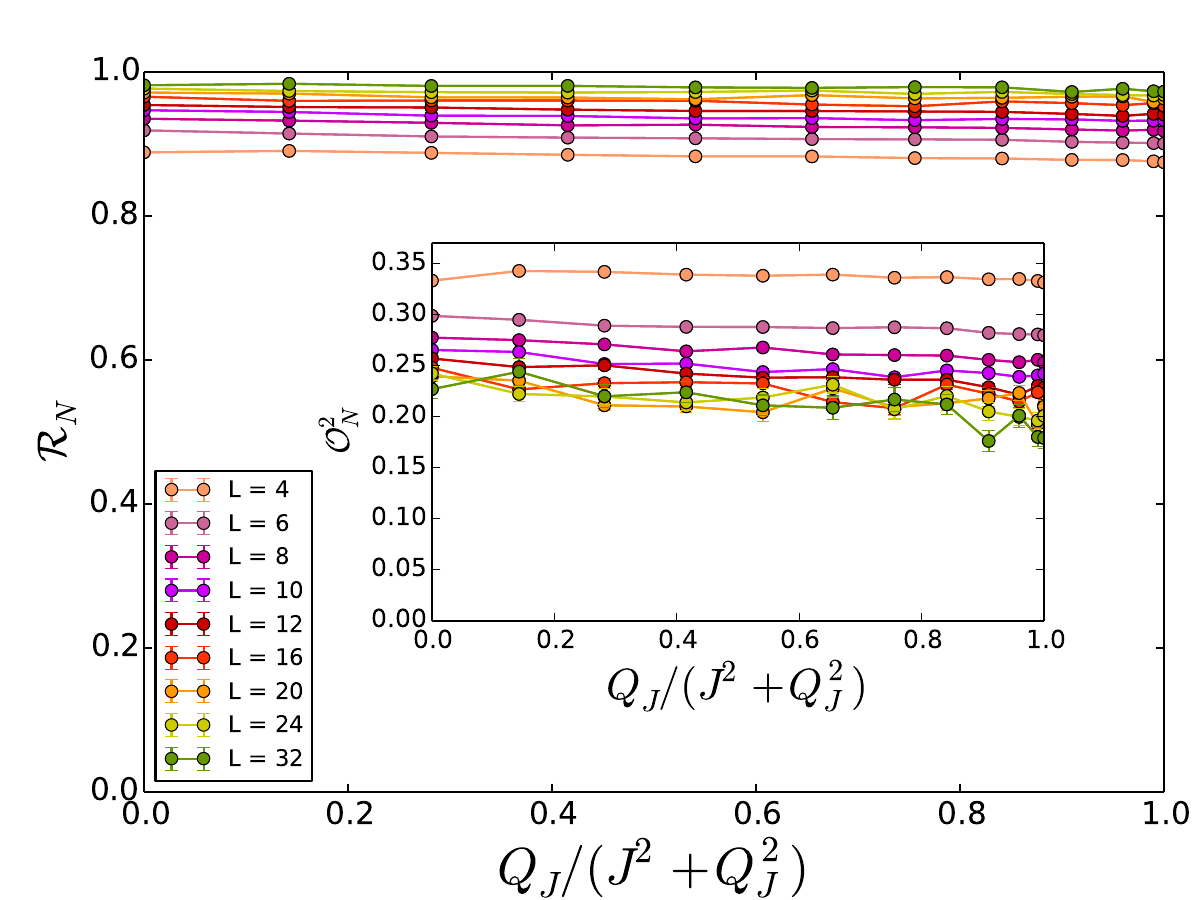}
\caption{Shown is the ratio $\mathcal{R}_{N}$ of the N\'eel order parameter of various values of the plaquette interaction coupling $Q_{J}$ with $J^{2} + Q_{J}^{2} = 1$ for systems 
of size $(L,L)$ with $L$ up to 32 lattice sites. $\mathcal{R}_{N}$ appears to be independent from $Q_{J}$ and approaches $1$ for increasingly large system sizes 
indicating a phase diagram consisting entirely of N\'eel order.
The inset shows the N\'eel order parameter $\mathcal{O}_{N}^{2}$.}\label{Fig_3}
\end{figure}

\begin{figure}[t]
\includegraphics[width=1.0\columnwidth]{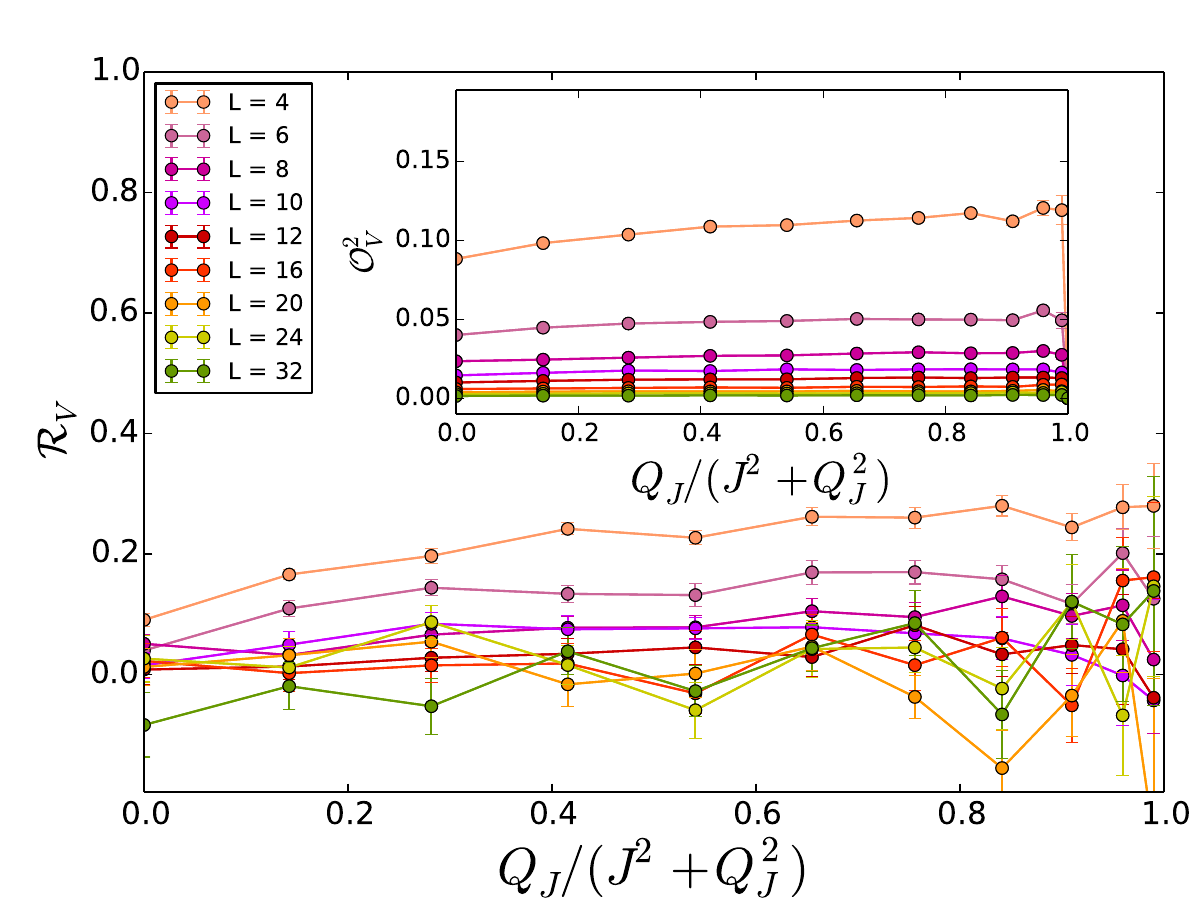}
\caption{Shown is the ratio $\mathcal{R}_{V}$ of the VBS order parameter of various values of the plaquette interaction coupling $Q_{J}$ with $J^{2} + Q_{J}^{2} = 1$ for systems  
of size $(L,L)$ with $L$ up to 32 lattice sites. Confirming the findings from Fig. \ref{Fig_3}, we see that $\mathcal{R}_{V}$ approaches zero for sufficiently large lattice sizes  
independent from the coupling $Q_{J}$ providing evidence for the absence of VBS order in the $J-Q_{J}$ model for spin$-1$.}\label{Fig_4}
\end{figure}

{\bf Measurements:}
In order to simplify the QMC loop algorithm, we have shifted our $J$ bond operators by the identity, $J(S_i \cdot S_j - 1)$.  The extensive energy quoted in the tables includes this shift.  In order to characterize the N\'eel and the VBS phases, we measure the equal time
bond-bond correlation function $\langle S_{\vec{r}}\cdot S_{\vec{r}+\hat{\alpha}} S_{\vec{r}^{'}}\cdot S_{\vec{r}^{'}+\hat{\alpha}} \rangle$. 
Here a bond is identified by its location on the lattice $\vec{r}$ and its
orientation $\alpha$ with $\alpha = x,y$ in two-dimensions.  In the VBS phase, lattice translational symmetry is broken. This gives rise 
to a Bragg peak in the Fourier transform of the bond-bond correlator defined as
\begin{equation}
\tilde{C}^{\alpha}(\vec{q})=\frac{1}{N_{\mathrm{site}}^2}\sum_{\vec{r},\vec{r}'}e^{i(\vec{r}-\vec{r}')\cdot\vec{q}}\langle
 S_{\vec{r}}\cdot S_{\vec{r}+\hat{\alpha}} S_{\vec{r}^{'}}\cdot S_{\vec{r}^{'}+\hat{\alpha}}\rangle \;.
\label{FT}
\end{equation}
For a columnar VBS patterns, peaks appear at the momenta $(\pi,0)$ and $(0,\pi)$ for $x$ and $y$-oriented bonds, respectively.  
Thus, the VBS order parameter is given by
\begin{equation}
\mathcal{O}_{VBS}=\frac{\tilde{C}^{x}(\pi,0)+\tilde{C}^{y}(0,\pi)}{2} \;.
\label{Ovbs}
\end{equation}

Another useful quantity to locate a possible phase transitions is the above mentioned VBS ratio $\mathcal{R}_{V}$. 
We first distinguish between $x-$ and $y-$oriented bonds: 
\begin{eqnarray}
\mathcal{R}^x_{V}&=&1- \tilde{C}^{x}(\pi,2\pi/L)/\tilde{C}^{x}(\pi,0) \nonumber \\
\mathcal{R}^y_{V}&=&1- \tilde{C}^{y}(2\pi/L,\pi)/\tilde{C}^{y}(0,\pi)\;.
\label{RXvbs}
\end{eqnarray}
Subsequently, we average over $x$ and $y$- orientations:
\begin{equation}
\mathcal{R}_{V}=\frac{\mathcal{R}^x_{V}+\mathcal{R}^y_{V}}{2} \;.
\label{RXvbs1}
\end{equation}
This quantity goes to $1$ in a phase with long-range VBS order and it approaches $0$ in a phase without VBS order present.   

The N\'eel structure factor is,
\begin{equation}
m^{2}_{z}(\vec{q})=\frac{1}{N_{\mathrm{site}}^2}\sum_{\vec{r},\vec{r}'}e^{i(\vec{r}-\vec{r}')\cdot\vec{q}}\langle
S^{z}_{\vec{r}}S^{z}_{\vec{r}'\alpha}\rangle \;.
\label{FT_m}
\end{equation}
The Bragg peak appears at momentum $(\pi,\pi)$ and thus the N\'eel order parameter is given by 
\begin{equation}
\mathcal{O}_{N}=m^{2}_{z}(\pi,\pi)\;.
\label{ONeel}
\end{equation}
To additionally provide a quantity that goes to $1$ in a N\'eel
ordered phase and vanishes in a phase without, we study the The N\'eel ratio:
\begin{eqnarray}
	\mathcal{R}^{x}_{N}&=&1- m^{2}_{z}(\pi+2\pi/L,\pi)/m^{2}_{z}(\pi,\pi) \nonumber \\
	\mathcal{R}^{y}_{N}&=&1- m^{2}_{z}(\pi,\pi+2\pi/L)/m^{2}_{z}(\pi,\pi)\;. 
\label{RN}
\end{eqnarray}
We can now average over both quantities:
\begin{equation}
\mathcal{R}_{N}=\frac{\mathcal{R}^x_{N}+\mathcal{R}^y_{N}}{2}\;.
\label{RXYNeel}
\end{equation}

The spin stiffness, $\rho_s$, is another quantity we use to detect the magnetic phase. It is defined as :
 \begin{equation}
  \rho_s=\frac{\partial^2 E(\phi)}{\partial \phi^2}\bigg|_{\phi=0} \\
  \label{eq:stiff}
 \end{equation}
Here E($\phi$) is the energy of the system when you add a twist of $\phi$ in the boundary condition in either the $x$ or the $y$ direction. In the QMC, this quantity is related to the winding number of loops in the direction that the twist has been added:
\begin{equation}
  \rho_s= \frac{\langle \mathcal{W}^2 \rangle}{\beta}\\
  \label{eq:winding}
 \end{equation}
 where $\beta$ is the inverse temperature. In the magnetic phase the stiffness extrapolates to a finite value in the thermodynamic limit, but goes to zero in the non-magnetic phase.

\begin{figure}
    \includegraphics[width=1.0\columnwidth]{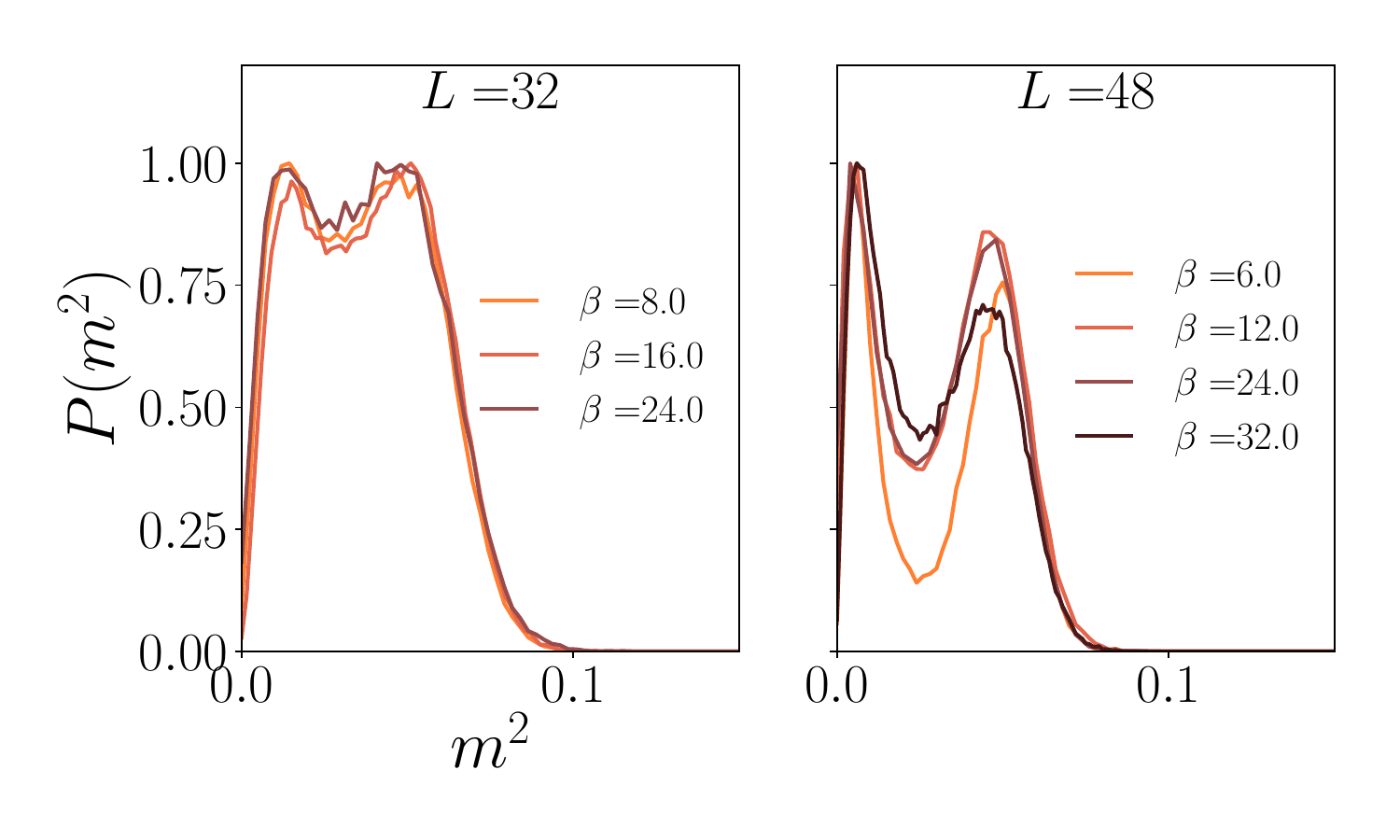}
    \caption{Magnetization histograms for $L=32$ (left) and $L=48$ (right) show two peaks near the transition, this feature is not significantly weakened on decreasing the temperature.}
    \label{fig:mag2histvsbeta}
\end{figure}

\subsection{Ground state convergence:}
\label{subsec:gsconv}
We investigate the behavior of
the observables described above in \ref{subsec:EDchecks} ($\mathcal{O}^2_V$, $\mathcal{O}^2_N$, $\rho_s$) when the SSE is carried out at different inverse
temperatures $\beta$. Fig. \ref{fig:obsvsbeta} shows that these quantities saturate as a function of inverse temperature $\beta$ before $\beta=6$. However close to the transition, one needs to go lower in temperature for saturation.
Therefore we do finite size scaling of histograms near the critical point for $\beta=L/4$ in order to probe the first order behavior. One can see from Fig. \ref{fig:mag2histvsbeta} that decreasing the temperature to $\beta>L/4$ does not significantly weaken the first order transition, so we can conclude that first order behavior persists at zero temperature.\\

\subsection{\texorpdfstring{$\mathbf{J-Q_J}$}{JQJ} Model for \texorpdfstring{$\mathbf{S=1}$}{S=1}}

\label{subsec:jqmodel}
We now briefly discuss another designer model Hamiltonian and compare the phase diagram for the two cases of a spin$-1/2$ system and a spin$-1$ system.

The so-called ``$J-Q$'' model was introduced by Sandvik in 2007 \cite{sandvik2007:deconf}. The model consists of a Heisenberg interaction
between nearest neighbor sites (see equation $(1)$ in the main manuscript) on the square lattice and an additional plaquette term:

\begin{eqnarray}\label{HQJ12}
        H_{Q} = -Q\sum_{ijkl \in \square} \big(\vec{S}_{i}\cdot \vec{S}_{j} - \frac{1}{4}\big)\big(\vec{S}_{k}\cdot \vec{S}_{l} - \frac{1}{4}\big)\;.
\end{eqnarray}
The spin$-1/2$ case of this model $H = H_{J} + H_{Q}$ was shown to have a phase transition from N\'eel order to VBS order at a critical point $ J/Q \approx 0.04$ \cite{sandvik2007:deconf}.

We now subject the same term structure to a SSE-MC simulation in order to determine the phase diagram. We note that for the spin$-1$ case the constant $\frac{1}{4}$    
is replaced by $1$ in order by make the plaquette term amenable to the SSE-MC study: 
\begin{eqnarray}
        H_{Q_{J}} = -Q_{J}\sum_{ijkl \in \square} \big(\vec{S}_{i}\cdot \vec{S}_{j} - 1\big)\big(\vec{S}_{k}\cdot \vec{S}_{l} - 1\big)\;.
\label{HQJ}
\end{eqnarray}
The $J-Q_{J}$ model spin$-1$ Hamiltonian is then $H_{JQ_{J}} = H_{J} + H_{Q_{J}}$. 
We analyzed the phase diagram for various couplings $J$ and $Q_{J}$ with the condition $J^{2} + Q_{J}^{2} = 1$ and found that the phase diagram consists 
entirely of N\'eel order independent from the ratio of the two coupling strengths $J$ and  $Q_{J}$. 
Fig. \ref{Fig_3} shows the ratio of the N\'eel order parameter. The ratio appears to be independent from $Q_{J}$ (with $J$ fixed by $J^{2} + Q_{J}^{2} = 1$). 
Further the ratio $\mathcal{R}_{N}$ approaches $1$ for increasingly large system sizes.
This is a clear indicator that the entire phase diagram consists of N\'eel order. For completeness we also give the ratio $\mathcal{R}_{V}$ of the VBS order parameter 
$\mathcal{O}_{V}^{2}$. In compliance with our findings from Fig. \ref{Fig_3}, we see the ratio $\mathcal{R}_{V}$ approaches zero for sufficiently large lattice sizes
independent from the coupling $Q_{J}$ (again with $J$ fixed by $J^{2} + Q_{J}^{2} = 1$). This provides evidence for the absence of VBS order that was present in the spin$-1/2$ 
flavor of the model.

\bibliography{career}

\end{document}